# Viewing citation trend of Indian physics and astronomy research papers since 2005 through the lens of some new indicators

Gopinath Das

Santal Bidroha Sardha Satabarshiki Mahavidyalaya,
Goaltore, Paschim Medinipur-721 128
West Bengal, India
e-mail: gopinathdas003@gmail.com

Bidyarthi Dutta
Dept. of Library and Information Science
Vidyasagar University
Midnapore 721 102, West Bengal, India
e-mail: bidyarthi.bhaswati@gmail.com

Anup Kumar Das
Centre for Studies in Science Policy
School of Social Sciences
Jawaharlal Nehru University
New Delhi 110 067, India
e-mail: anup_csp@mail.jnu.ac.in

## Abstract

The indicator *Citation Swing Factor (CSF)* has recently been developed to quantitatively measure the diffusion process from the h-core zone to the h-core excess zone. This paper calculated CSF for Indian physics and astronomy research output appeared in selective Indian journals from 2005 to 2020. The theoretical values of CSF are also calculated based on its fundamental equation and the same was compared with the respective observed values. The average error over the entire period is found 2.26% indicating proximity between theoretically expected and practically observed values. Besides, three other scientometric indicators are introduced here, viz. *Time-normalized Total Cited Ratio (TC)*, *Time-Normalised Cited Uncited Ratio (CU)*, and *Time-Normalised Total Uncited Ratio (TU)*. Of these four indicators, the variation of TC is highest (1.76), followed by TU (0.53), CU (0.37), and CSF(E) (0.09), as evident from the values of respective Coefficients of Variations. The numerical values of these indicators are found out for the same sample and the temporal variations along with their mutual interrelationships are determined by regression analysis. It is observed that the three indicators, TC, CU, and TU are mutually interrelated through the following linear regression equations: $TC = -0.76 + 1.88*TU$ and $CU = 0.201 + 0.34*TU$

## Keywords

h Index; Excess Citation; e Index; R Index; Total Citation; h-core Citation; Citation Diffusion; Citation Swing Factor; Cited to Uncited Ratio;

# 1. Introduction

The term 'Citation' implies a connection between a part or whole of the cited document and a part or whole of the citing document, broadly known as source document[1]. A 'Reference' is the acknowledgment that one document gives to another and a 'Citation' is the acknowledgment that one document receives from another[2]. According to Garfield[3], there are many reasons behind the existence of citation. The citation analysis is the most popular technique used in scientometrics that helps in evaluating the quality of research publications, assessing the contribution of authors and the standard of journals. Eugene Garfield[4-8] illustrated in several articles the potentialities of citation analysis in the evaluation of research faculty. According to Price[9], citation patterns in research articles indicate the research front in a particular subject domain. The citation is a recognition of intellectual works that are reckoned as principal rewards of science[10]. This paper has verified the observed values of a recently introduced indicator, viz. *Citation Swing Factor*, with its formula-based theoretically calculated values for a sample of Scopus-indexed 18357 Indian physics and astronomy research publications from 2005 to 2020, which received 91245 citations. Besides, this paper proposed three citation-based indicators, calculated their values for the same sample, and determined the inter-relationship among them.

# 2. Literature Review

Citation trend analysis includes the study of changing the number of citations received by articles or journals over the years. Usually, the articles or journals of a particular discipline or subject are studied in this context. The citation trends of journals devoted to subjects like primatology[11], psychological medicine[12], forensic science[13], and behavioral psychology[14] were analyzed. The citation trend analysis of physical therapy research output was done by Imai[15] et al. Morgan[16] investigated whether citation trends reflect epidemiologic patterns. Gazni[17] analyzed journal self-citation trends and Ajibade & Stephen[18] analyzed citation trends on E-government in South African countries. Giovanni[19], Biradar & Kumbar[20], Chi[21], and Singh[22] analyzed citation trends on psychiatry, environmental science, political science, and defence science respectively. The bibliometric indicators developed before 2k evaluated Journals based on citation count and the number of papers. The concept of author-level indicators and article-level indicators were developed after 2k. The introduction of the h index by Hirsch[23] in 2005 was the milestone of modern or post-2k metrics. A scientist has an h-index equal to H if the top H of his/her N publications from a ranked list have at least H citations each[23]. Besides, there are numbers of indices developed so far known as h-type indices[24].

The citation trend of physics research output was observed by Alvarez, Vanz & Barbosa[25], where analysis of Brazilian research on High Energy Physics from 1983 to 2013 was incorporated. Scientometric indicators for output, collaboration, and impact studies were used to characterize the field. Tsay[26] carried out a comparative study between scientometric data including the number of source items, the number of citations, impact factor, immediacy index, citing half-life and cited half-life, for crucial journals in physics, chemistry, and engineering. Mugnaini, Packer & Meneghini[27] compared the average h-index of the members of the Brazilian Academy of Sciences with the members of the National Academy of Sciences (USA) for 10 different areas of science. Czerwon[28] analyzed the dynamics of the subject domain, i.e. theoretical high energy physics based on six-year periods' (1979-1984) citations. This paper provided clues to understanding the growth of a new research domain from a core body of seminal literature. Mohan & Kumbar[29] carried out scientometric studies of the publications on stellar and galactic astrophysics research in India during the last 20 years.

Makhoba & Pouris[30] analyzed Activity and Attractivity indices of South African research output and compared the same with BRICS nations in biotechnology, energy, astronomy, and paleontology from 2002 to 2012. Li et al[31] carried out a bibliometric analysis of publications in the journal *Symmetry* from 2009 to 2019. Using bibliometric data generated through a model of citation dynamics, Medo & Cimini[32] compared several indicators for the scientific impact of individual researchers of physics. Andre[33] analyzed 1.2 million research articles on LASER published since 1960 to present some bibliometric studies and found the h-index of 590. Moed & Raan[34] developed bibliometric indicators for researchers in physics and astrophysics based on citation-per-publication ratio and researchers' individual perception. Flores, Raga & Roy[35] scientometrically evaluated articles published from 2010 to 2019 contributed by Mexican astronomers. Henneken & Kurtz[36] developed bibliometric indicators based on the number of citations, number of reads, and number of downloads of the articles. Wildegaard[37] calculated 17 author-level indicators for 512 researchers in Astronomy, Environmental Science, Philosophy, and Public Health. Havemann & Larsen[38] tested 16 bibliometric indicators concerning their validity for the individual astrophysics researcher by estimating their power to predict later successful researchers.

## 3. Research Gap

The literature review shows enough works on citation trend analysis in different subject fields. A substantive number of analytical studies on author-level and institution-level bibliometric indicator development have also been observed. But, except for one article[31], no work is found discussed on indicator analysis for the physics and astronomy journals. Also, only one work[29] is observed in the Indian context dealing with indicator analysis in the Indian astrophysics research domain. The absence of studies on journal-level bibliometric indicator analysis particularly in the Indian context has created a research gap in this domain. It is historically justified that physics is the field where Indian contributions during both pre-and post-independence eras have been outstanding. It is borne out by the fact that one physicist from pre-independent India received the Noble Prize, and scientists like J.C. Bose, M.N. Saha, S.N. Bose, and K.S. Krishnan missed it narrowly. There are several esteemed physics and astronomy journals that started in colonial India and still continuing. It is thus imperative to carry out journal-level bibliometric indicator analysis in the Indian physics and astronomy research domain. This paper has calculated and analytically interpreted four indicators (CSF, TC, CU & TU) for 15 core Indian physics and astronomy journals. The indicator CSF has recently been formulated[39] and the other three indicators are introduced here.

## 4. New Citation-based Indicators

### *4.1 Citation Swing Factor (CSF)*

One of the major objectives of h-type indices was to normalize the h-index by dividing the same by the number of publications or the age of citation (time normalization). An author or journal just after receiving one citation enters the domain of the cited vs. citing graph (Figure 1) through the tail zone, which is the entry point for a cited item. The number of citations received may be increased in due course of time causing the said cited item to gradually shifting from the tail zone towards the h-core zone and eventually the h-excess zone. This continuous movement of a cited item in the cited-citing graph (Figure 1) with the accretion of citation may be termed as diffusion of cited item. The indicator *Citation Swing Factor* (CSF) has recently been developed to measure this diffusion quantitatively[39], which is the ratio of change in FHE ($d\theta$) to change in FET ($d\varepsilon$). The parameters FHE and FET indicate the fraction of h-core to excess citations and fraction of excess to total citations respectively and may be defined as FET ($\varepsilon$) = $\sqrt{\dfrac{E}{T}}$ and FHE ($\theta$) = $\sqrt{\dfrac{H}{E}}$, where T, H, and E stand for Total Citation, h-Core Citation and Net Excess Citation respectively. The ($\dfrac{d\theta}{d\varepsilon}$) is thus equivalent to the fractional h-core to excess citations with respect to fractional excess citations over total citations. The observed or experimental value of CSF that is followed from the basic definition is represented as ($\dfrac{d\theta}{d\varepsilon}$). Here both θ and ε are continuous variables.

The differentiation of θ with respect to ε yielded the value ($\dfrac{-R^3}{he^2}$), where $R^2$, $h^2$ and $e^2$ indicate total citations, h-core citations, and excess citations respectively. The indicator CSF thus points out the shift of h-core citations with respect to the fold of excess citations to total citations, which in turn, figures the citation shift from h-core to the h-excess zone.

$$\text{CSF} = \frac{d\theta}{d\varepsilon} = \frac{-R^3}{he^2} \quad \text{.......(1)}$$

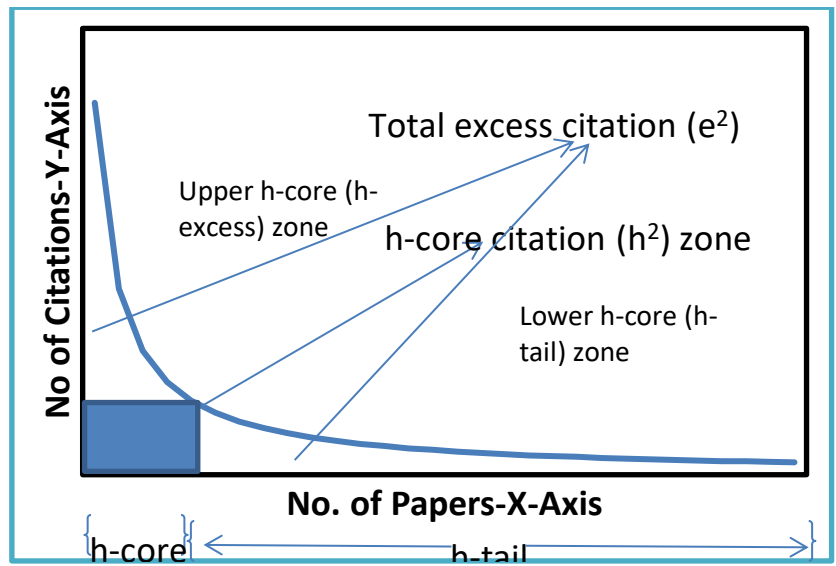

Figure 1: Three h-zones (excess, core, and tail) in a cited vs. citing graph

### 4.2. Time-normalized Total Cited Ratio (TC)

Let '$n$' number of articles belonging to any subject '$S$', was published in an arbitrary year '$y_1$', of which '$k$' number of articles altogether received '$c$' citations in any later year '$y_2$' ($y_2 > y_1$). Hence, the $(n-k)$ number of articles remained uncited in the year $y_2$. The Time-Normalised Total Cited Ratio, denoted by TC is defined as the ratio between the total number of published articles ($n$) to the number of cited articles ($k$), divided by $(y_2 - y_1)$. The difference between two years, $(y_2 - y_1)$ may be regarded as the age of the publication.

$$TC = \frac{n}{k(y_2 - y_1)} \quad \ldots\ldots(2)$$

The TC is a measure of the fold of the total number of published articles with respect to the total number of cited articles. It figures out the relative abundance of the total number of articles with respect to cited articles per unit age of the publication.

### 4.3. Time-Normalised Cited Uncited Ratio (CU)

The Time-Normalised Cited Uncited Ratio, denoted by CU is defined as the ratio between the number of cited articles (k) to the number of uncited articles $(n-k)$, divided by $(y_2 - y_1)$. The CU is a measure of the fraction of cited articles with respect to the total number of uncited articles. It figures out the relative strength of the cited articles with respect to the uncited articles per unit age of the publication.

$$CU = \frac{k}{(n-k)(y_2 - y_1)} \quad \ldots\ldots(3)$$

### 4.4. Time-normalized Total Uncited Ratio (TU)

The Time-Normalised Total Uncited Ratio, denoted by TU is defined as the ratio between the number of total published articles ($n$) to the number of uncited articles $(n-k)$, divided by $(y_2 - y_1)$. The TU is a measure of the fold of total published articles with respect to the number of uncited articles. It figures out the relative fraction of the uncited articles with respect to the total number of published articles per unit age of the publication.

$$TU = \frac{n}{(n-k)(y_2 - y_1)} \quad \ldots\ldots(4)$$

## 5. Research Question

1) Is the observed values of CSF (represented by $\frac{d\theta}{d\varepsilon}$) tally with the theoretical values of the same (represented by $\frac{-R^3}{he^2}$) for Indian physics and astronomy research output from 2005 to 2020 as represented in equation (1)?

2) How do the values of the indicators $TC$, $CU$ and $TU$ for Indian physics and astronomy research output from 2005 to 2020 change with time or publication age?

3) In there any inter-relationship exists among these three indicators, i.e. $TC$, $CU$ and $TU$?

## 6. Scope and Methodology

The values of the new indicators viz. CSF, TC, CU, and TU are calculated for Indian physics and astronomy research output appeared in fifteen esteemed Indian physics journals from 2005 to 2020 (Table 1 (CSF) & Table 2 (TC, CU & TU)). The observed values of CSF are calculated based on available data and the same have been compared with the respective theoretical values. Of the fifteen journals, eight journals belong to the core domain of physics and astronomy (S. No. 1, 3, 5, 6, 7, 8, 10, and 13), while five journals belong to allied interdisciplinary areas of physics but publish articles on physics regularly (S. No. 2, 4, 9, 11 and 12). The last two journals belong to the entire natural science discipline but publish physics articles on regular basis. These two journals are very old and esteemed Indian science journals. The list of the said fifteen esteemed Indian journals selected for this study is furnished below:

1) Bulletin of the Astronomical Society of India
2) Defence Science Journal
3) Indian Journal of Biochemistry and Biophysics
4) Indian Journal of Engineering and Materials Sciences
5) Indian Journal of Physics
6) Indian Journal of Pure & Applied Physics
7) Indian Journal of Radio and Space Physics
8) Journal of Astrophysics and Astronomy
9) Journal of Earth System Science
10) Journal of Medical Physics
11) Journal of Scientific and Industrial Research
12) Journal of Vibrational Engineering and Technologies
13) Pramana - Journal of Physics
14) Proceedings of the Indian National Science Academy
15) Proceedings of the National Academy of Sciences India Section A - Physical Sciences

The necessary data for calculating these four indicators have been collected from the *Scopus* database. The search strategy followed in *Scopus* under 'Advanced Search' was, "SUBJAREA(PHYS) AND AFFILCOUNTRY (INDIA) AND (EXACTSRCTITLE(BULLETIN OF THE ASTRONOMICAL SOCIETY OF INDIA)). The time range was set from 2005 to 2020. The same strategy was repeated for the other fourteen journals as listed above and all fifteen results were summed up at last, which resulted 18357 articles in total. The total number of citations, h-core citations, and excess citations figured 91245, 12361, and 78884 respectively. The year-wise breakup of the data is presented in Table 1. Based on yearly figures of h-core and h-excess citations, FET and FHE are calculated. The consecutive annual changes in the values of FHE and FET yielded $d\theta$ and $d\varepsilon$ respectively. The ratio of $d\theta$ to $d\varepsilon$ or $\frac{d\theta}{d\varepsilon}$ gives the observed value of CSF, which is compared with the theoretical value, i.e. ($\frac{-R^3}{he^2}$), where $R^2$, $h^2$, and $e^2$ indicate total citations, h-core citations, and net excess citations respectively. The total number of cited and uncited papers over the stipulated period figured 12757 and 5600 respectively. The yearly figures of the total, cited and uncited articles (Table 1) yielded TC, CU, and TU (Table 2) from 2005 to 2020. The temporal variations of CSF, TC, CU, and TU are observed. The correlation and regression analyses between TC-CU, CU-TU, and TC-TU have been carried out to delineate their mutual interrelationships.

## 7. Results and Analysis

The Scopus database indexed 18357 research publications on physics and astronomy from 2005 to 2020 appeared in fifteen above-listed journals, which received 91245 citations. The average number of citations per research publication over this period figures 5. The year-wise break-up of total citations ($R^2$), h-core citations ($h^2$), and net excess citations ($e^2$) are presented in Table 1, to calculate FET and FHE. The changes between successive years' FET and FHE values are represented by $d\varepsilon$ and $d\theta$ respectively. The observed value of CSF, i.e. CSF(O) is calculated by dividing $d\theta$ by $d\varepsilon$, which is represented by $\frac{d\theta}{d\varepsilon}$, and the expected or theoretical value of CSF, i.e. CSF(E) is given by equation (1). It is found from Table 1 and Figure 2, that the observed values are in close proximity with expected values with an average error of 2.26%, which asserts the validity of equation (1) for Indian physics and astronomy research output from 2005 to 2020. The temporal variations of the magnitudes of CSF(O) [dotted line] and CSF(E) [solid line] are presented in Figure (2), which shows the average, maximum and minimum values of CSF as 3.3, 3.8, and 2.9 respectively. The standard deviation of 15 observed and 15 expected values of CSF altogether is 0.278 or approximately 28%. The negative Kurtosis values of both CSF (Table 3) indicate their flat distribution with a thin tail that accords near-constancy of this indicator over the said period. The coefficients of variation and standard deviations of CSF are fairly low, i.e. less than 0.1 and 1 respectively, which also signals its near-constancy.

Out of 18357 publications, 12757 (≈70%) articles received the citation(s) while 5600 (≈30%) articles remained uncited to date. The year-wise break-up of total (n), cited (k), uncited (n-k) publications, and age of publications ($y_2-y_1$) are given in Table 2, where $y_2$ indicates the current year, i.e. 2021 and $y_2$ indicates publication years ranging from 2005 to 2020. The values of the indicators TC, CU, and TU are calculated following the equations (2), (3), and (4) respectively, and presented in Table 2. The variations of TC, CU, and TU with publication age are presented in Figure 3. Of these four indicators, the variation of TC is highest (1.76), followed by TU (0.53), CU (0.37), and CSF(E) (0.09), as evident from the values of respective Coefficients of Variations (CV) (Table 3). The standard deviation is highest for TC (0.74), followed by TU (0.33), CSF (0.27), and CU (0.15) (Table 3). The graph for TC is highly skewed and possesses a thick tail that is endorsed by the high Kurtosis value of TC (Table 3). The variation of TC with publication age follows Harris Model as best fit graph with the equation, $TC = \dfrac{1}{a + by^c}$ ……… (5), where, $y = y_2 - y_1$ =Publication Age, $a = -5.789$, $b = 6.114$ & $c = 0.242$. The graph for TU is a little bit skewed and possesses a thin tail that is endorsed by the low Kurtosis value of TU (Table 3). The variation of TU with publication age follows Rational Function as best fit graph with the equation, $TU = \dfrac{a + by}{1 + cy + dy^2}$ …….(6), where, $y = y_2 - y_1$ = Publication Age, $a = 7.55*10^{(-13)}$, $b = 4.346*10^{(10)}$, $c = 2.404*10^{(10)}$ and $d = 7.068*10^{(10)}$.

The low kurtosis and standard deviations of TU indicate the poor variability or comparatively stronger constancy of TU. The graph of CU (Figure 3) shows an almost constant pattern that is also endorsed by its low coefficient of variation and standard deviation values. The negative Kurtosis values indicate its flat distribution with a thin tail revealing constancy. The Correlation Coefficient between TC and CU is found as 0.26, indicating a weak positive correlation. The Correlation Coefficients between CU-TU and TC-TU are 0.74 and 0.84 respectively indicating strong positive correlations. The linear regression equations of CU on TU & of TC on TU are found out as: $CU = 0.201 + 0.34*TU$ ……(7) (Coefficient of Determination ($R^2$) = 0.55; Standard Error = 0.11) and $TC = -0.76 + 1.88*TU$ …….(8) (Coefficient of Determination ($R^2$) = 0.70; Standard Error = 0.42).

Table 1: Temporal variations of CSF since 2005 to 2020

| | 2005 | 2006 | 2007 | 2008 | 2009 | 2010 | 2011 | 2012 | 2013 | 2014 | 2015 | 2016 | 2017 | 2018 | 2019 | 2020 |
|---|---|---|---|---|---|---|---|---|---|---|---|---|---|---|---|---|
| T | 6910 | 6157 | 6002 | 7224 | 7697 | 8428 | 7705 | 7471 | 7842 | 6814 | 4869 | 4318 | 3807 | 2499 | 1925 | 1577 |
| H | 1156 | 1024 | 1024 | 1296 | 1296 | 1156 | 961 | 841 | 1089 | 625 | 484 | 361 | 400 | 256 | 196 | 196 |
| E | 5754 | 5133 | 4978 | 5928 | 6401 | 7272 | 6744 | 6630 | 6753 | 6189 | 4385 | 3957 | 3407 | 2243 | 1729 | 1381 |
| FET ($\varepsilon$) | 0.91 | 0.91 | 0.91 | 0.91 | 0.91 | 0.93 | 0.94 | 0.94 | 0.93 | 0.95 | 0.95 | 0.96 | 0.95 | 0.95 | 0.95 | 0.94 |
| FHE ($\theta$) | 0.45 | 0.45 | 0.45 | 0.47 | 0.45 | 0.40 | 0.38 | 0.36 | 0.40 | 0.32 | 0.33 | 0.30 | 0.34 | 0.34 | 0.34 | 0.38 |

| | | | | | | | | | | | | | | | |
|---|---|---|---|---|---|---|---|---|---|---|---|---|---|---|---|
| $d\varepsilon$ | | 0.001 | -0.002 | -0.005 | 0.01 | 0.02 | 0.01 | 0.01 | -0.01 | 0.03 | -0.004 | 0.01 | -0.01 | 0.001 | 0.0003 | -0.012 |
| $d\theta$ | | -0.002 | 0.01 | 0.01 | -0.02 | -0.05 | -0.02 | -0.02 | 0.05 | -0.08 | 0.01 | -0.03 | 0.04 | -0.005 | -0.001 | 0.040 |
| CSF (O) | | -2.939 | -2.930 | -2.898 | -2.903 | -3.023 | -3.181 | -3.295 | -3.229 | -3.343 | -3.577 | -3.642 | -3.600 | -3.464 | -3.485 | -3.357 |
| CSF (E) | | -2.936 | -2.941 | -2.919 | -2.877 | -2.930 | -3.129 | -3.235 | -3.359 | -3.116 | -3.635 | -3.522 | -3.774 | -3.447 | -3.481 | -3.489 |
| % Error | | 0.09 | 0.38 | 0.73 | 0.90 | 3.15 | 1.64 | 1.85 | 3.85 | 7.28 | 1.60 | 3.40 | 4.62 | 0.49 | 0.12 | 3.80 |

Table 2: Temporal variations of TC, CU, and TU from 2005 to 2020

| | 2005 | 2006 | 2007 | 2008 | 2009 | 2010 | 2011 | 2012 | 2013 | 2014 | 2015 | 2016 | 2017 | 2018 | 2019 | 2020 |
|---|---|---|---|---|---|---|---|---|---|---|---|---|---|---|---|---|
| $n$ | 947 | 935 | 903 | 1072 | 974 | 1041 | 1148 | 1144 | 1065 | 1319 | 1185 | 1224 | 1164 | 1120 | 1414 | 1702 |
| $k$ | 754 | 685 | 660 | 803 | 791 | 852 | 866 | 894 | 897 | 993 | 870 | 903 | 831 | 702 | 700 | 556 |
| $\dfrac{n}{k}$ | 1.26 | 1.36 | 1.37 | 1.33 | 1.23 | 1.22 | 1.33 | 1.28 | 1.19 | 1.33 | 1.36 | 1.36 | 1.40 | 1.60 | 2.02 | 3.06 |
| $(y_2 - y_1)$ | 16 | 15 | 14 | 13 | 12 | 11 | 10 | 9 | 8 | 7 | 6 | 5 | 4 | 3 | 2 | 1 |
| TC | 0.078 | 0.091 | 0.098 | 0.103 | 0.103 | 0.111 | 0.133 | 0.142 | 0.148 | 0.19 | 0.227 | 0.271 | 0.35 | 0.532 | 1.01 | 3.061 |
| $(n - k)$ | 193 | 250 | 243 | 269 | 183 | 189 | 282 | 250 | 168 | 326 | 315 | 321 | 333 | 418 | 714 | 1146 |
| $\dfrac{k}{n-k}$ | 3.91 | 2.74 | 2.72 | 2.99 | 4.32 | 4.51 | 3.07 | 3.58 | 5.34 | 3.05 | 2.76 | 2.81 | 2.50 | 1.68 | 0.98 | 0.49 |
| CU | 0.244 | 0.183 | 0.194 | 0.230 | 0.360 | 0.410 | 0.307 | 0.397 | 0.667 | 0.435 | 0.460 | 0.563 | 0.624 | 0.560 | 0.490 | 0.485 |
| $\dfrac{n}{n-k}$ | 4.91 | 3.74 | 3.72 | 3.99 | 5.32 | 5.51 | 4.07 | 4.58 | 6.34 | 4.05 | 3.76 | 3.81 | 3.50 | 2.68 | 1.98 | 1.49 |
| TU | 0.307 | 0.249 | 0.265 | 0.307 | 0.444 | 0.501 | 0.407 | 0.508 | 0.792 | 0.578 | 0.627 | 0.763 | 0.874 | 0.893 | 0.990 | 1.485 |

Table 3: Statistical parameters of the indicators' values over the entire time span

| Parameters / Indicators | Mean | Median | Range | Standard Deviation | Coefficient of Variation | Kurtosis |
|---|---|---|---|---|---|---|
| CSF(O) | 3.26 | 3.29 | 0.74 | 0.27 | 0.08 | -1.47 |
| CSF(E) | 3.25 | 3.24 | 0.90 | 0.3 | 0.09 | -1.36 |
| TC | 0.42 | 0.15 | 2.98 | 0.74 | 1.76 | 12.25 |
| CU | 0.41 | 0.42 | 0.48 | 0.15 | 0.37 | -0.99 |
| TU | 0.62 | 0.54 | 1.24 | 0.33 | 0.53 | 1.61 |

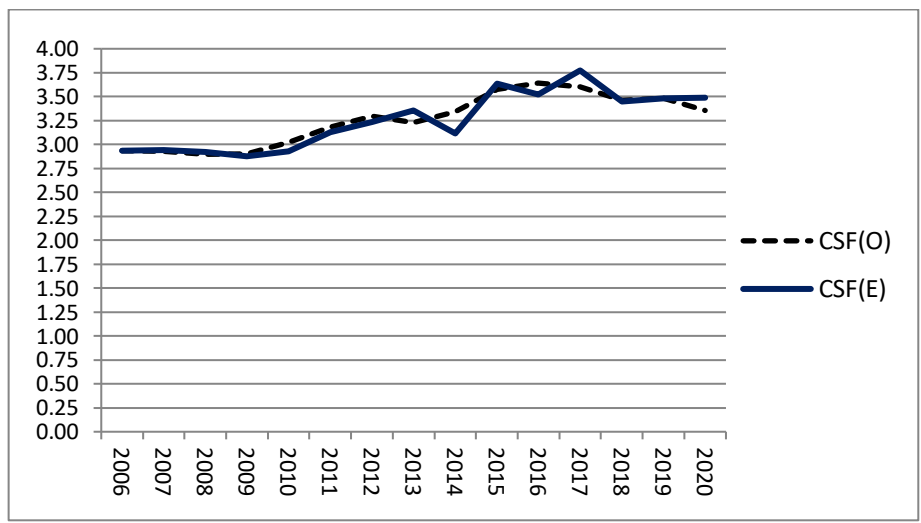

**Figure 2: Temporal variation of CSF(O) and CSF(E)**

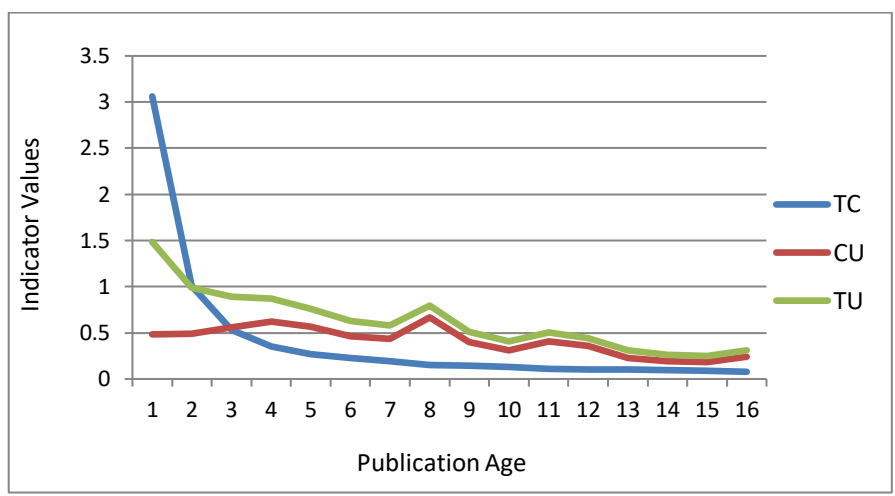

**Figure 3: Variation of TC, CU & TU with Publication-Age**

## 8. Conclusion

In this paper, the theoretical value of the 'Citation Swing Factor' (CSF) is compared with its observed values that show close propinquity pointing out the appropriateness of the theory for the sample of this study. It needs to be tested for other samples from other subject domains. The CSF measures the relative share of h-core citations with respect to net excess citations. Its variation indicates the shift of h-core citations towards the h-excess zone. The almost constancy of CSF values over 15 years indicates the steadiness of the diffusion process of h-core citations to the h-excess zone. The majority of articles remained long in the h-core zone and shifted slowly to the h-excess zone. The near constancy of CSF values over 15 years indicates a steady citation accumulation rate of Indian physics and astronomy articles. But, the CSF has its limitations, because the total citation, as well as h-core citation, are manipulable through self-citation, re-citation, or coercive citation. Hence it is important to develop modified CSF excluding the manipulated citations owing to biased and corrupted practices. The modified CSF is supposed to portray the true picture of citation diffusion.

The indicators TC, CU, and TU are time normalized, i.e. the respective actual values are divided by the respective age of publications. This is done to minimize the bias arisen due to the time dependence of the citation accumulation process. The citation accumulation is accelerated with the increase in publications' age. The decreasing pattern of TC and TU with age proves that the number of cited articles increases with the age of publication. The variation of TU however, is very trifle. The constancy of CU reveals that the number of articles and number of cited articles both hike at an almost equal pace. The TU is linearly related with both CU and TC (equations (7) and (8)). The changing patterns of TC, CU, and TU need to be studied for other subject domains also to analyze the growth pattern of cited articles with respect to total and uncited articles. These indicators include the number of cited and uncited articles along with the total number of articles. But, in any discipline, quite a large number of articles remain once cited or twice cited only. Hence, except 'cited' and 'uncited' articles, another category of articles may be introduced, i.e. 'poorly cited' articles to modify the formalism of TC, CU, and TU.

## 9. Acknowledgment